\documentclass[onecolumn,showpacs,amsmath,amssymb]{revtex4}
\usepackage{graphicx,epsf,subfigure}               
\usepackage{dcolumn}                
\usepackage{bm}                     
\usepackage{epstopdf}
\usepackage{lineno}

\newcommand{\vect}[1]{{\bf #1}}

\newcommand{\eqi}{\begin{equation}}
\newcommand{\eqo}{\end{equation}}

\begin{document}

\title{Libration driven elliptical instability}

\author{D. C\'ebron$^{1,2}$}
\email{david.cebron@erdw.ethz.ch}
\author{M. Le Bars$^1$}
\author{J. Noir$^{2,3}$}
\author{J.M. Aurnou$^3$}

\affiliation{
$^1$ CNRS and Aix-Marseille Univ., IRPHE (UMR 7342), 13384, Marseille cedex 13, France.\\
$^2$ Institut f\"ur Geophysik, ETH Z\"urich, Sonneggstrasse 5, Z\"urich, CH-8092, Switzerland. \\
$^3$ Department of Earth and Space Sciences, University of
California, Los Angeles, CA 90095-1567  USA.}

\date{\today}

\begin{abstract}

The elliptical instability is a generic instability which takes
place in any rotating flow whose streamlines are elliptically
deformed. Up to now, it has been widely studied in the case of a constant, non-zero differential rotation between the fluid and
the elliptical distortion with applications in turbulence,
aeronautics, planetology and astrophysics. In this letter, we extend
previous analytical studies and report the first numerical and
experimental evidence that elliptical instability can also be driven
by libration, i.e. periodic oscillations of the differential
rotation between the fluid and the elliptical distortion, with a
zero mean value. Our results suggest that intermittent,
space-filling turbulence due to this instability can exist in the liquid cores and sub-surface oceans of so-called synchronized planets and
moons.
\end{abstract}



\maketitle


The longitudinal libration of a so-called synchronized planet or
moon, i.e. the oscillation of its axial rotation rate whose mean
value is otherwise equal to the orbital rotation rate, arises
through its gravitational coupling with its closest neighbors
\cite[][]{yoder95,comstock03}. In the body interior, the interaction
of a fluid layer (e.g. an iron rich liquid core or a subsurface
ocean) with the surrounding librating solid shell resulting from
viscous, topographic, gravitational or electromagnetic coupling,
leads to energy dissipation and angular momentum transfer that need to be accounted for in the planet thermal history and orbital
dynamics, and possibly in its magnetic state \cite[][]{LeBars2011}. A number of studies has been devoted to
libration-driven flows in axisymmetric containers in order to
investigate the role of the viscous coupling. It has been shown that
longitudinal libration in an axisymmetric container can drive
inertial waves in the bulk of the fluid as well as boundary layer
centrifugal instabilities in the form of Taylor-G\"ortler rolls
\cite[][]{aldridgephd,
aldridge69,aldridge75,tilgner99,noir09,calkins2010lib}. In addition,
laboratory and numerical studies
\cite[][]{wang1970,calkins2010lib,noir2010zf_a,sauret2010} have
corroborated the analytically predicted generation of a mainly
retrograde axisymmetric and stationary zonal flow in the bulk, based
upon non-linear interactions within the Ekman boundary layers
\cite[][]{wang1970,busse2010zf_b,busse2010zf_a,calkins2010lib,noir2010zf_a}.
Although practical to isolate the effect of viscous coupling, the
spherical approximation of the core-mantle or ice shell-subsurface
ocean boundaries, herein generically called the CMB, is not fully
accurate from a planetary point of view and very restrictive from a
fluid dynamics standpoint. Indeed, due to the rotation of the
planet, the gravitational interactions with companion bodies and the low order spin-orbit resonance of the librating planets we
are considering, the general figure of the CMB must be ellipsoidal
with a polar flattening and a tidal bulge pointing on average toward
the main gravitational partner \cite[][]{Goldreich2010}.

The fluid dynamics that occurs in a rapidly rotating ellipsoidal
cavity has been widely studied in the case of constant but different
rotation rates of the fluid and the elliptical distortion. This
corresponds in geophysical terms to a non-synchronized body with a
constant spin rate $\Omega_0$, subject to dynamical tides rotating at
the constant orbital rotation rate $\Omega_{orb}$. In particular, it
has been shown that this elliptically deformed base flow can be destabilized
by the
so-called tidally-driven elliptical instability or TDEI
\cite[][]{kerswell02, Cebron2011}.

Generally speaking, the elliptical instability can be seen as the inherent
local instability of elliptical streamlines \cite[][]
{bayly1986three,waleffe1990three,Ledizes2000}, or as the parametric resonance between two free inertial waves (resp. modes) of the rotating
unbounded (resp. bounded) fluid and an elliptical strain, which is not an inertial wave or mode  \cite[][]
{moore1975instability,tsai1976stability}. Such a resonance mechanism,
confirmed by numerous works in elliptically deformed cylinders \cite[e.g.][]
{eloy2000experimental,eloy2001stability,eloy2003elliptic} and ellipsoids
\cite[][]
{lacaze04,lacaze2005elliptical,LeBars2010}, also operates for triadic resonance of three inertial modes, proposed to explain the secondary instability of the elliptical instability \cite[][]
{mason1999nonlinear,kerswell1999secondary}. This triadic resonance of three inertial modes also applies for the
inertial precessional instability in cylinders \cite[][]
{lagrange2008instability,lagrange2011precessional} and spheroids \cite[][]
{kerswell1993instability,wu2009dynamo}: there, the base flow forced by
precession is itself an inertial mode (e.g. the so-called Poincar\'e or
tilt-over mode in the spheroid
\cite{Poincare_precession,busse1968steady,noir2003experimental,cebron2010tilt}),
which resonates with two inertial modes.

It has been shown that selected resonances of the TDEI are sensitive to the ratio of the rotation rates of the fluid and the elliptical
distortion \cite{malkus1989,
LeBars2010}. In particular, the elliptical instability is known to vanish in the case of synchronous rotation $\Omega_0 = \Omega_{orb}$. But
theoretical arguments suggest that oscillations around this
synchronous state may be sufficient to excite elliptical instability
\cite[][]{kerswell98,herreman2009,Cebron2011}. This could be of
fundamental importance in planetary liquid cores and subsurface
oceans of synchronized bodies, where librations are generically
present. This letter thus aims at validating the existence of a
libration driven elliptical instability, hereafter referred to as
LDEI. To do so, we first extend previous analytical studies
\cite[][]{kerswell98,herreman2009,Cebron2011} using a local WKB
(Wentzel-Kramers-Brillouin)
approach that allows us to determine a generic formula for the
growth rate of LDEI. We then present the numerical and experimental
validation of the existence of the LDEI, in good agreement with the
theoretical results. Finally, implications for planets and moons are
briefly discussed.

\begin{figure}
  \begin{center}
    \begin{tabular}{lr}
      \subfigure[]{\includegraphics[scale=0.44]{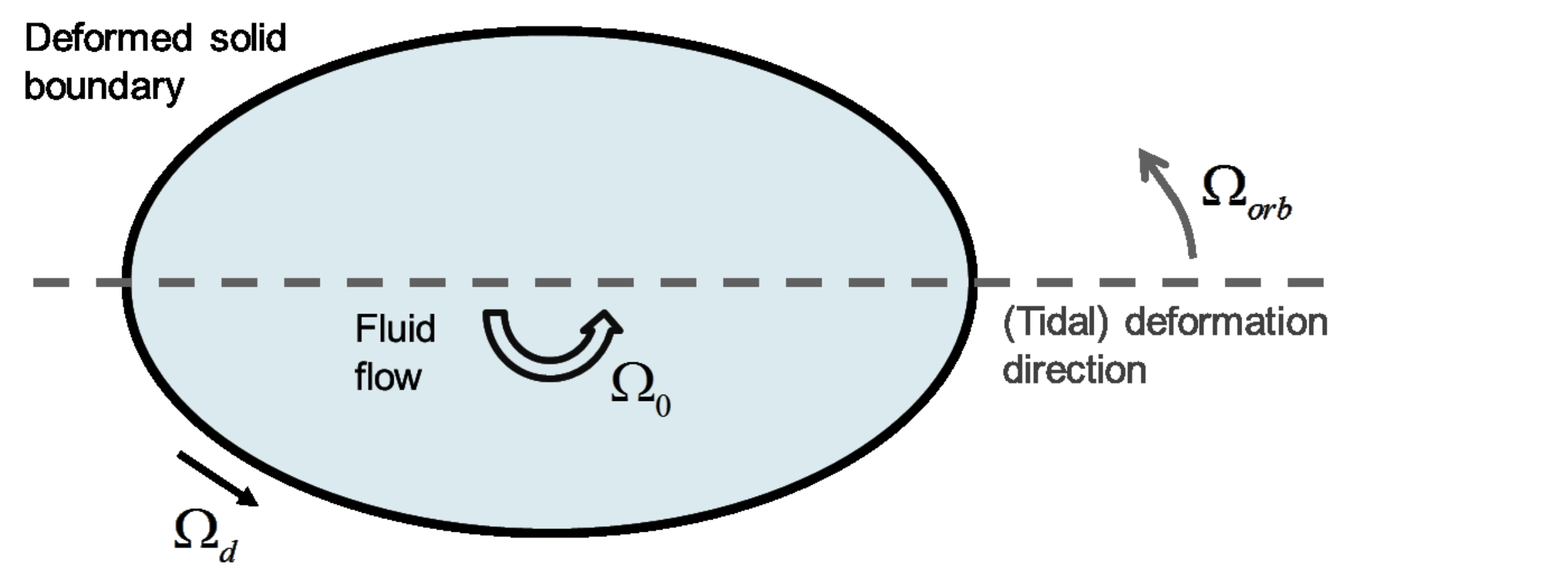}} &
      \subfigure[]{\includegraphics[scale=0.36]{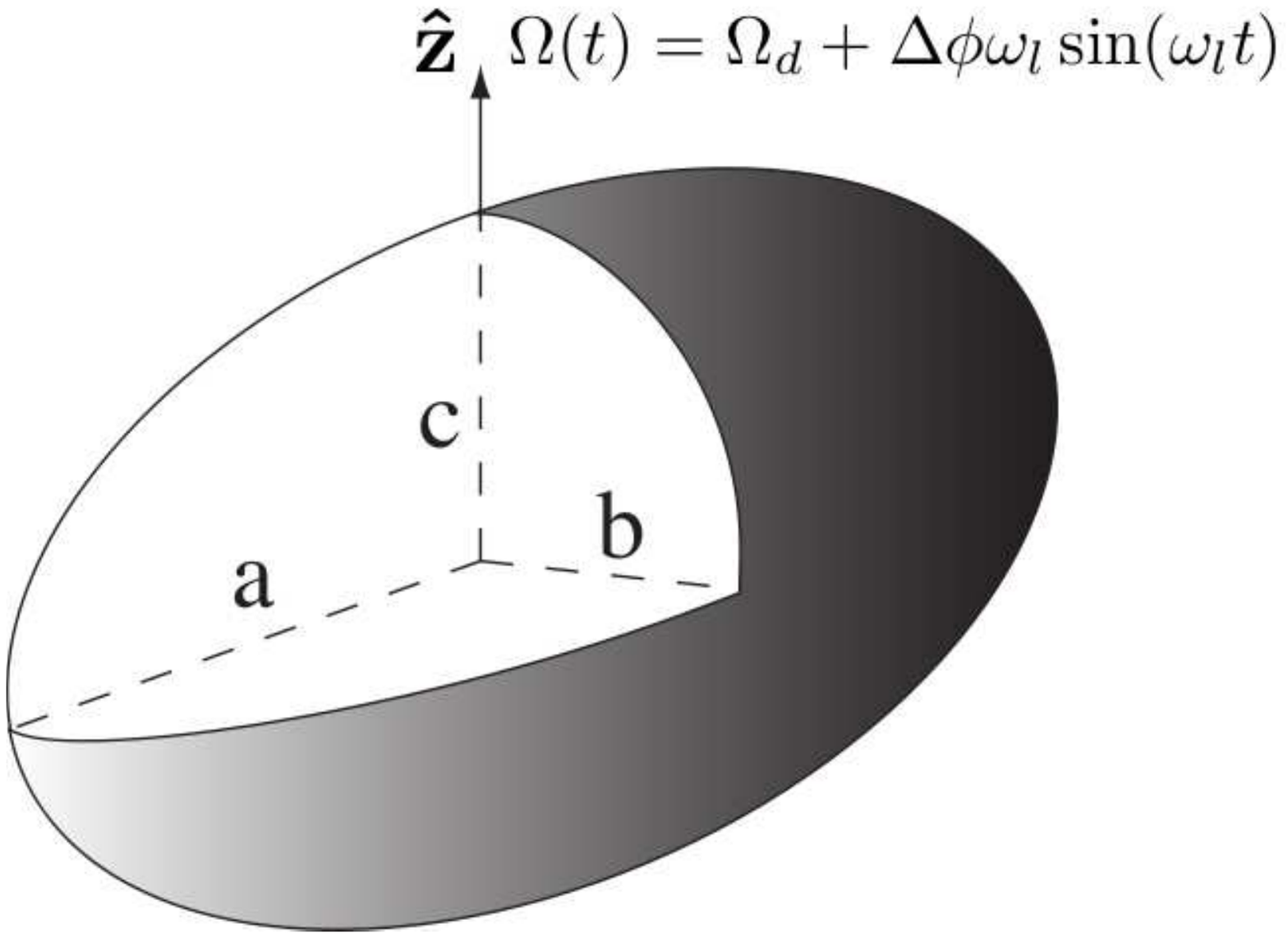}} 
     \end{tabular}
    \caption{\it{(a) Mean rotation rates involved in our modelized planetary fluid layer: the elliptical (tidal) deformation rotates at the mean orbital rate $\Omega_{orb}$, the external solid boundary has a constant tangential velocity imposed by the mean planetary spin rate $\Omega_d$ and the fluid mean rotation rate in the bulk of the fluid is $\Omega_0$ (possibly different from $\Omega_d$ in a general case \cite{sauret2010}). (b) Schematic view of the oscillating triaxial ellipsoidal container. }}
    \label{oblong_spheroid}
  \end{center}
\end{figure}

We consider a homogeneous, non-conductive, incompressible, newtonian
fluid enclosed in a librating triaxial ellipsoid (fig.\ref{oblong_spheroid}). In the container frame of reference, the equation of the ellipsoidal boundary is written as
$x^2/a^2+y^2/b^2+z^2/c^2=1$,  where ($x,\,y,\,z$) is a cartesian
coordinate system with its origin at the center of the ellipsoid,
with $\vect{\hat{x}}$ along the long equatorial axis $a$,
$\vect{\hat{y}}$ along the short equatorial axis $b$, and
$\vect{\hat{z}}$ along the rotation axis $c$. We define the
ellipticity $\beta=(a^2-b^2)/(a^2+b^2) $ and the aspect ratio
$c^*=c/R, $ where $R=\sqrt{(a^2+b^2)/2}$ is the mean equatorial radius. The motion of longitudinal libration of the
container is modeled by a time dependency of its axial rotation
rate $ \Omega(t)=\Omega_d + \Delta\phi \, \omega_l \sin (\omega_l
t)$, where $\Omega_d$ is the mean rotation rate of the container ($d$ for diurnal), $\Delta\phi $
the amplitude of libration in radians and $\omega_l$ the angular
frequency of libration. In the frame of reference attached to the
container, the equations of motion, made dimensionless using $R$ as
the length scale and $\Omega_d^{-1}$ as the time scale, become:
\begin{eqnarray}\label{NS}
\frac{\partial \vect{u}}{\partial t}-\vect{u}\times\left( \nabla \times \vect{u} \right) + 2\ [1+\varepsilon \sin (ft)]\ \vect{\hat{z}}\times \vect{u}&=&-\nabla \Pi + E\ \nabla^2 \vect{u} -\varepsilon f \cos (ft) (\vect{\hat{z}}\times\vect{r}),\\
\nabla \cdot \vect{u}&=&0.\label{NS2}
\end{eqnarray}
In (\ref{NS}), $\Pi$ is the reduced pressure, which includes the
time-variable centrifugal acceleration. The Ekman number $E$ is defined by $E=\nu/(\Omega_d R^2)$, where $\nu$ is the kinematic
viscosity, the dimensionless libration frequency $f$ is defined as
$f=\omega_l/\Omega_d$, and $\varepsilon$ is the libration forcing
parameter defined by $\varepsilon=\Delta\phi f$. In the limit of small Ekman numbers, the flow can be
decomposed into an inviscid component $\vect{U}$ in the volume and a
viscous boundary layer flow $\tilde{\vect{u}}$ that satisfies the no-slip boundary condition on the CMB.
Introducing this separation,
\cite{kerswell98} proposed the following solution to the inviscid
equations of motion subject to the non-penetration condition at the
CMB:
\begin{eqnarray}
\vect{U}&=&-\varepsilon \sin (ft) \left[
\vect{\hat{z}}\times\vect{r}-\beta\ \nabla (xy)
\right].\label{sol_inv1}
\end{eqnarray}

It can be shown that eq. (\ref{sol_inv1}) provides a solution of the
inviscid form of (\ref{NS}) in the bulk; the inertial forces are
balanced by the pressure gradient. No net zonal flow can
result from the non-linear interactions in the quasi-inviscid
interior \cite[][]{busse2010zf_b}, since pressure gradients alone cannot drive net zonal flows.  However, the no-slip
boundary condition is not entirely fulfilled by this solution:
viscous corrections in the Ekman boundary layer must be considered,
whose non-linear interactions generate a zonal flow in the bulk
\cite[][]{wang1970,busse2010zf_a,busse2010zf_b}, as observed in
axisymmetric containers
\cite[][]{wang1970,calkins2010lib,noir2010zf_a,sauret2010}.


In addition to these laminar and mostly two-dimensional (2D) motions, Kerswell and Malkus \cite{kerswell98} first suggested that turbulent three-dimensional (3D) motions can be excited by an elliptical instability
corresponding to a parametric resonance involving two free inertial
waves of the rotating fluid and strain of the inviscid base flow
(\ref{sol_inv1}), from which energy is extracted. Since the base
flow is of azimuthal wavenumber $m=2$ and temporal frequency $f$,
this parametric resonance occurs only when $ m_a-m_b=\pm m =\pm 2 \quad
\mbox{and} \quad \lambda_a-\lambda_b=\pm f$, where $m_a, m_b$ are
the azimuthal wavenumbers of the two resonating free inertial waves
and $\lambda_a, \lambda_b$ are their frequencies non-dimensionalized
by $\Omega_d$. In addition to these resonance conditions, the two
waves must have close radial and azimuthal structures to interact
positively, corresponding to the so-called principal resonances
\cite{Eloy2000}. This set of rules forms the basis for a global
analysis of the elliptical instability, which thus requires an exact
description of the inertial modes in the considered ellipsoidal
geometry. Unfortunately, little is known about inertial modes for the finite values of $\beta$ considered in the present study. This makes a complete characterization of the
elliptical instability by global analysis presently out of reach  for our simulations and
experiments. 

As an alternative, a local
stability analysis of the base flow (\ref{sol_inv1}) can be
performed, independently of the geometry of the container. Here, we
make use of the results presented in \cite{herreman2009} for the
special case $f=1$, later complemented \cite{Cebron2011}. The
local analysis is based on the WKB method
\cite[][]{Ledizes2000}, which allows an upper bound to be
derived for the growth rate of the elliptical instability. In this
approach, perturbations are assumed to be sufficiently localized so
as to be advected along flow trajectories. The perturbations are
sought in the form of local plane waves characterized by their
wavevector $\mathbf{k}(t)$, with norm $k \gg 1$, and tilted by an
angle $\zeta$ to the rotation axis. Elliptical instability appears
by resonance of two identical plane waves, only differing by their
direction of propagation. The inviscid growth rate,
$\sigma_{inv}$, can be determined by solving the Euler
equations at the first order in the forcing amplitude $\beta
\varepsilon$, yielding 

\eqi \sigma_{inv} =
\frac{16+f_{res}^2}{64}\beta \varepsilon, \eqo where $f_{res}\neq 0$
stands for a resonant forcing frequency \cite[see details
in ref.][]{Cebron2011}. In the strict WBK limit $k \gg 1$, the dispersion
relation between the forcing frequency and the excited plane waves
is $f_{res}/2=\pm 2 \cos{\zeta}$; hence, all forcing frequencies
between $-4$ and $+4$ should be resonant. However, by accounting for
the shape of the container and for viscous damping, similarly to the TDEI, resonances are only possible for selected couples of inertial waves,
especially at the rather large Ekman numbers accessible to numerics \cite[][]{Ledizes2000,LeBars2010}.  Thus, the system resonates only for selected values of the
forcing frequency. Introducing a small detuning between the
libration frequency $f$ and a given exact resonance $f_{res}$
\cite[see method in ref.][for the standard case of TDEI]{Ledizes2000} and
taking into account the dominant viscous damping in the Ekman layer,
it can be shown that excitation of the instability takes place
around each resonant frequency in a band $f \in
[f_{res}-\sigma_{inv} ; f_{res}+\sigma_{inv}]$, where the
typical growth rate is \eqi \sigma = \sqrt{\sigma_{inv}^2 -
(f_{res}-f)^2} - K E^{1/2}, \label{sigmath} \eqo 
with $K \in [1;10]$ a constant of order $1$, which depends on the excited mode. The first
term on the left hand side of (\ref{sigmath}) defines the range of
unstable frequencies around a particular resonance; the second term
describes the viscous damping of the instability. Besides the strict
WBK limit, we expect this equation to be generally valid, once the
specific values of the resonant frequency $f_{res}$ and of the
damping factor $K$ have been determined \cite[see validation for
the case of TDEI in ref.][]{LeBars2010}. Since both $f_{res}$ and $K$
depend on the selected inertial waves, both values can vary with the shape of the container.




\begin{figure}
  \begin{center}
    \includegraphics[scale=0.75]{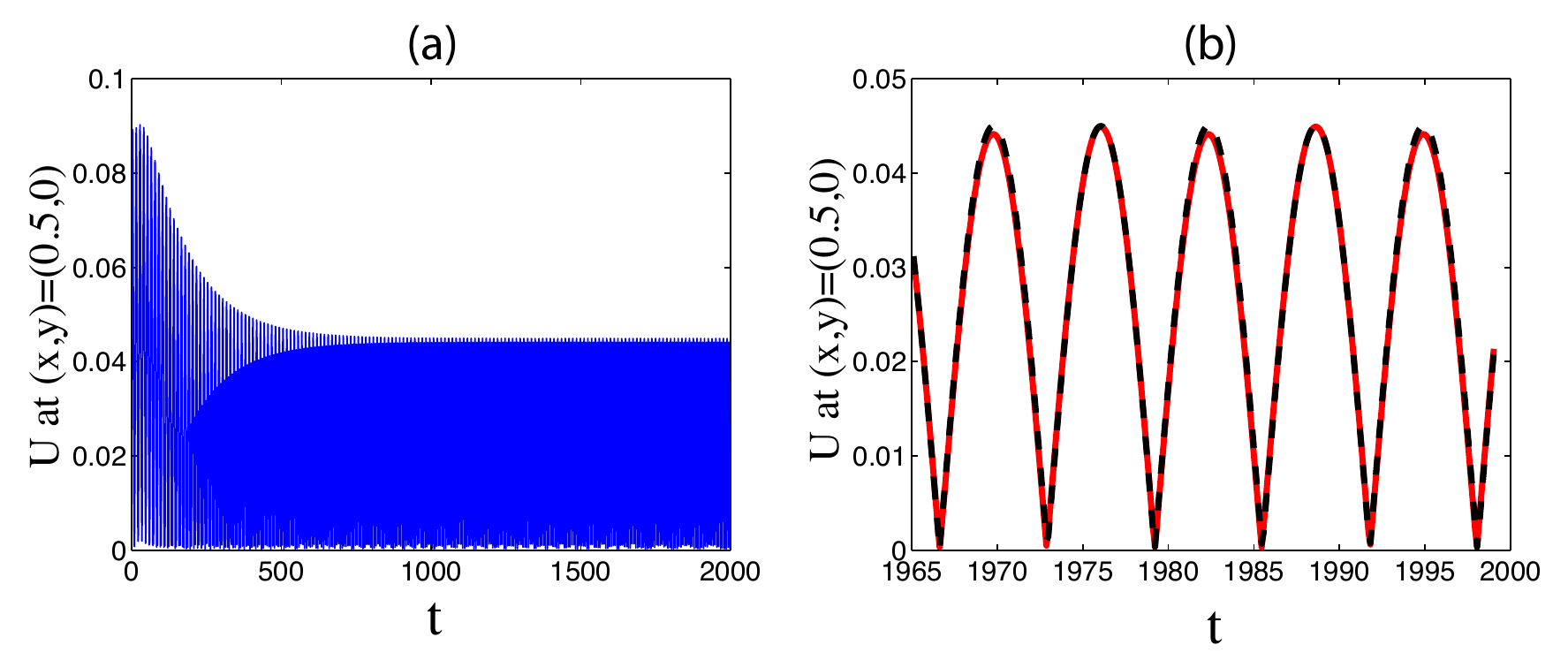}
    \caption{\it{(color online)
    (a) Temporal evolution of $U=|\vect{u}|$ at the location $(x=0.5,y=0)$, as calculated
by the 2D version of our numerical model for $f=0.5$,
$\varepsilon=0.1$, $\beta=0.1$ and $E=4\times10^{-4}$. (b) Zoom ($t>1965$) of figure
(a) once the flow is established: the continuous gray curve (red online) stands
for the numerical results, and the dashed black curve for the
theoretical flow (\ref{sol_inv1}).}}
    \label{David1}
  \end{center}
\end{figure}
%
%

\begin{figure}
  \begin{center}
    \leavevmode
    \includegraphics[scale=0.4]{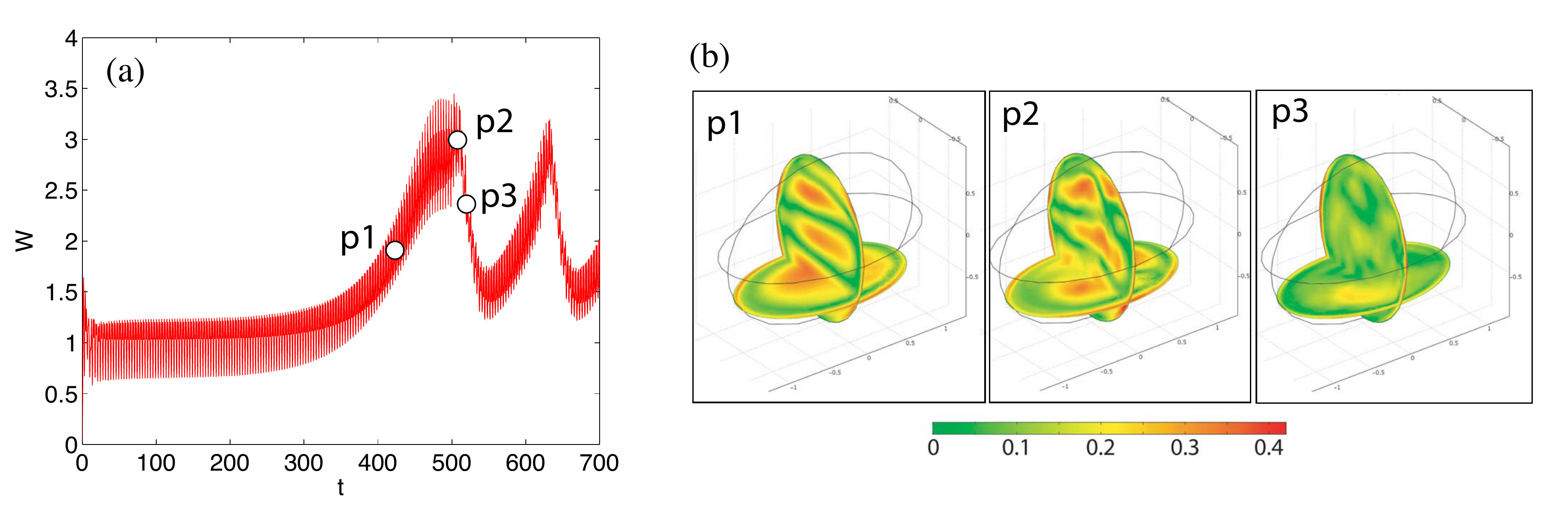}
    \caption{\it{(color online) (a) Time evolution of the absolute value of
the axial velocity integrated over the whole container,
non-dimensionalized by the mean value between times $t=100$ and
$t=200$ (i.e. after the spin-up and before the potential
destabilization). Numerical simulations are performed for $E=5\times
10^{-4}$, $\beta=0.44$, $f=1.76$, $\varepsilon=0.92$ and $c^*=0.95$. (b) $|\vect{u}|$, in
a meridional cross section and in the plane $z=-0.5$. The sequence shows, from left to right, the typical field $|\vect{u}|$ during the exponential growth, at saturation and during
the collapse.}}
    \label{David4}
  \end{center}
\end{figure}
To quantitatively validate the existence of LDEI and the above
analytical results, we combine a systematic numerical study with
selected laboratory experiments. We use the software
Comsol Multiphysics\textsuperscript{\textregistered}  based on the finite elements method to perform
our simulations. We work in the container
frame of reference where the ellipsoidal shape of the container is
stationary, and we solve the equations of motion (\ref{NS}) subject to no-slip boundary conditions, starting from a fluid at rest at
time $t=0$. We refer the reader to \cite{Cebron2010} for more
informations on the numerical method. A first series of computations
has been performed in 2D to test the realization of the inviscid
base flow (\ref{sol_inv1}). A typical result is shown in figure
\ref{David1}. After a transient behavior scaling as a
typical viscous time in $t \sim E^{-1}$ (figure \ref{David1}a), the
theoretical base flow (\ref{sol_inv1}) is indeed established in the
bulk (figure \ref{David1}b), whereas the corrections necessary to
fulfill the tangential part of the boundary conditions are
restricted to the Ekman layer of depth $E^{-1/2}$. In the 3D ellipsoid, the Ekman
pumping associated with the Ekman layer acts to significantly accelerate the establishment of the base
flow, and we thus expect (\ref{sol_inv1}) to be the starting
velocity field. 

Series of numerical simulations have then been
performed in a triaxial ellipsoid as a function of the libration
frequency $f$, the libration amplitude $\varepsilon$ and the aspect
ratio $c^*$, keeping $E=5\times 10^{-4}$ and $\beta=0.44$. An
example of the temporal evolution of the absolute value of the
axial velocity integrated over the volume, $W$, is presented in
figure \ref{David4}a for the case $f=1.76$, $\varepsilon=0.92$ and
$c^*=0.95$. Libration driven elliptical instability is present, characterized by intense 3D motions with rich temporal dynamics on typical times much longer than the spin and libration
periods. In particular, we observe characteristic cycles of growth,
saturation, collapses and relaminarization towards the base flow,
already observed for the classical case of TDEI
\cite[][]{LeBars2010}. The typical changes in the flow field
during one cycle are illustrated in figure \ref{David4}b which shows
the norm of the velocity in meridional and equatorial cross
sections. 

\begin{figure}
  \begin{center}
    \begin{tabular}{ccc}
      \subfigure[]{\includegraphics[scale=0.43]{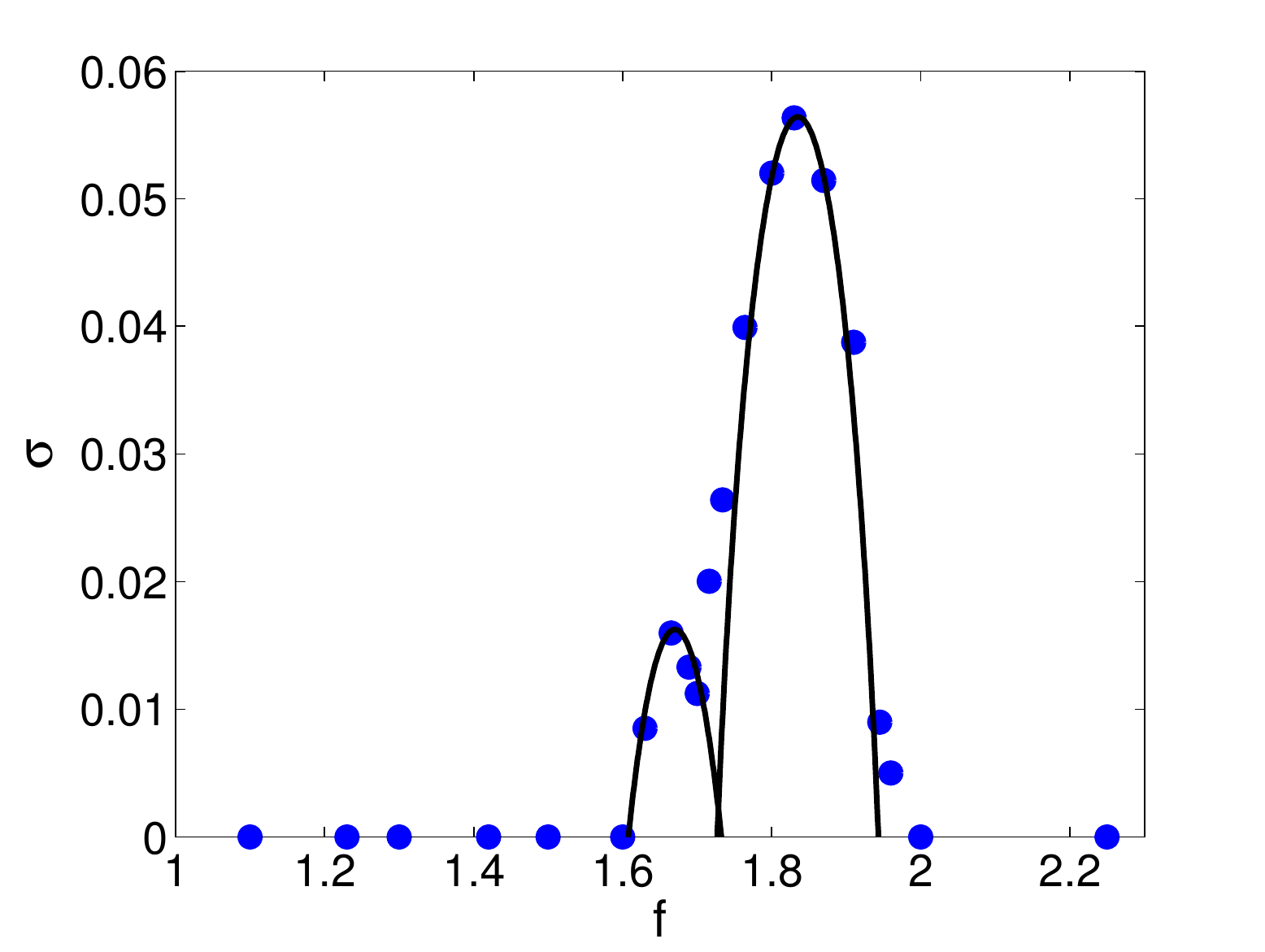}} &
      \subfigure[]{\includegraphics[scale=0.46]{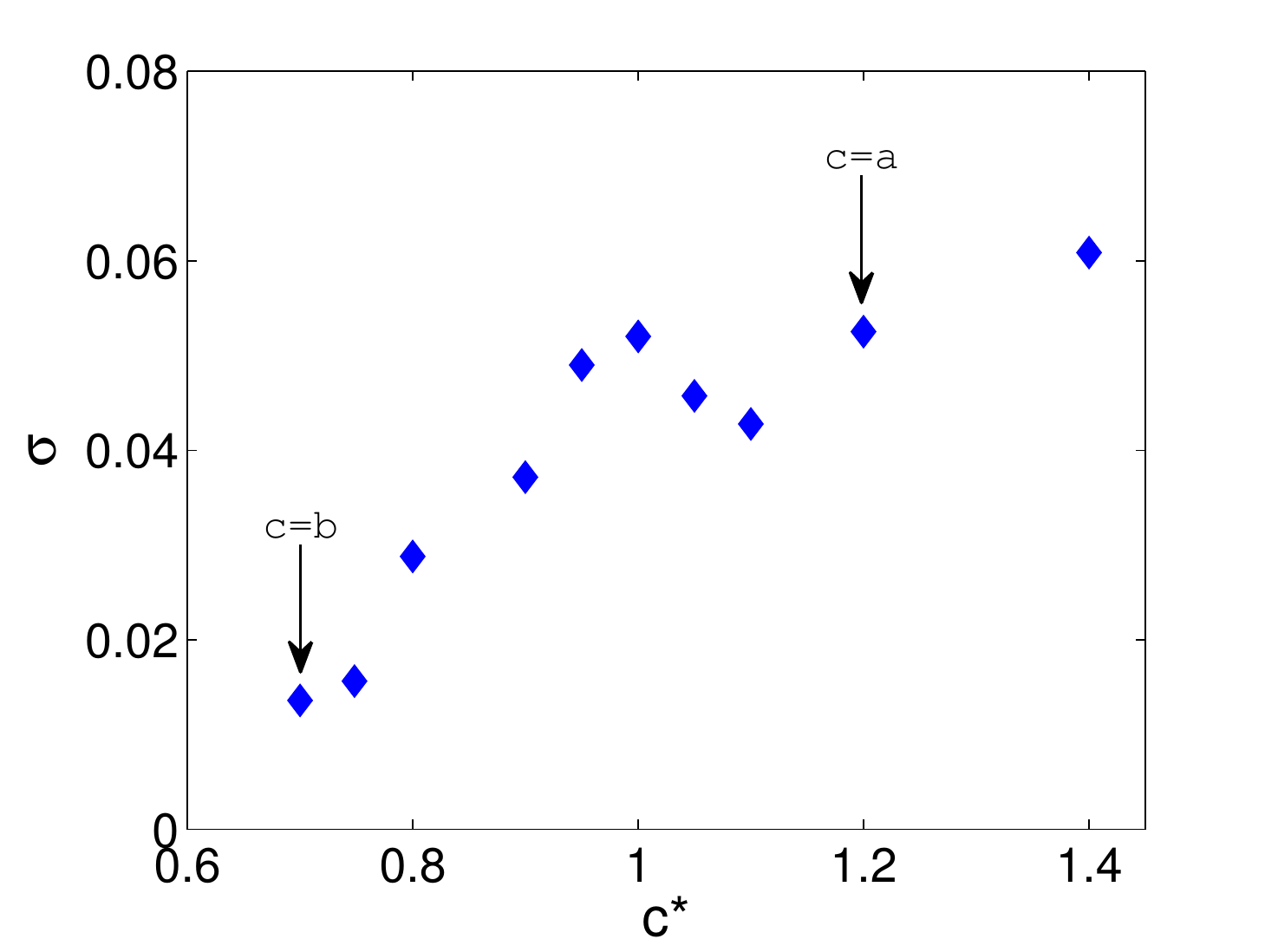}}
     \end{tabular}
    \caption{\it{(color online) Systematic numerical determination of
the growth rate $\sigma$ of the LDEI for $E=5\times 10^{-4}$,
$\beta=0.44$, $\varepsilon=0.991$ ($\sigma$ is set to $0$ when the LDEI is not excited). (a) As a function of the
libration frequency $f$ for $c^*=1$. Also shown here as a
solid curve is the theoretical growth rate (\ref{sigmath}), where $f_{res}$ and $K$ have been determined by adjusting the analytical formula to each of the two local maxima of
the numerical growth rate. (b) As a function of the aspect ratio $c^*$ for $f=1.8$. Note that the LDEI is excited in spheroidal geometries ($c=b$ or $c=a$).
}}
    \label{LDEI_freq}
  \end{center}
\end{figure}

The growth rate of LDEI can be obtained by fitting the
growing parts of time series of the integrated axial velocity $W$
with an exponential function. The systematic evolution of this
growth rate with $f$ for $c^*=1$ is shown in figure
\ref{LDEI_freq}a, in comparison with the analytical formula
(\ref{sigmath}), where $f_{res}$ and $K$ have been determined by
adjusting (\ref{sigmath}) to each of the two local maxima of the
numerically determined growth rate. We observe two bands of
frequency centered around $f_{res}\sim1.83$ and $f_{res}\sim1.67$
where an elliptical instability grows. Good agreement is recovered for all neighboring values of $f$, validating
the generic equation (\ref{sigmath}). 

The systematic evolution of the growth rate as a function of the aspect ratio $c^*$ at a fixed
resonant frequency $f=1.8$ is shown in figure \ref{LDEI_freq}b.
Although less dramatically than the frequency detuning, the geometrical
factor $c^*$ can also alter the growth rate of the
elliptical instability at a given frequency. We observe that the
optimal geometry at the resonant frequency $f=1.83$ is realized for
$c^*\sim 1$, as already shown in \cite{Cebron2010} for the classical
elliptical instability (TDEI). In contrast with the TDEI studied in \cite{Cebron2010}, we have no theoretical arguments that precludes the excitation of the LDEI in spheroidal geometries ($c=b$ or $c=a$). In figure \ref{LDEI_freq}b, we label these two particular cases, showing that the growth rates in these geometries are positive. Note that this does not refute the conclusions of \cite{zhang10}, who show that libration in longitude cannot produce a direct resonance of an inertial mode. Here, the resonance does not occur directly between a mode and the forcing but between two modes and the forcing in a parametric coupling. The modes excited in the LDEI are not at the frequency of the forcing (which would be the case for a direct forcing), it is their frequency difference that is equal to $f$. 

\begin{figure}
  \begin{center}
    \leavevmode
    \includegraphics[scale=0.59]{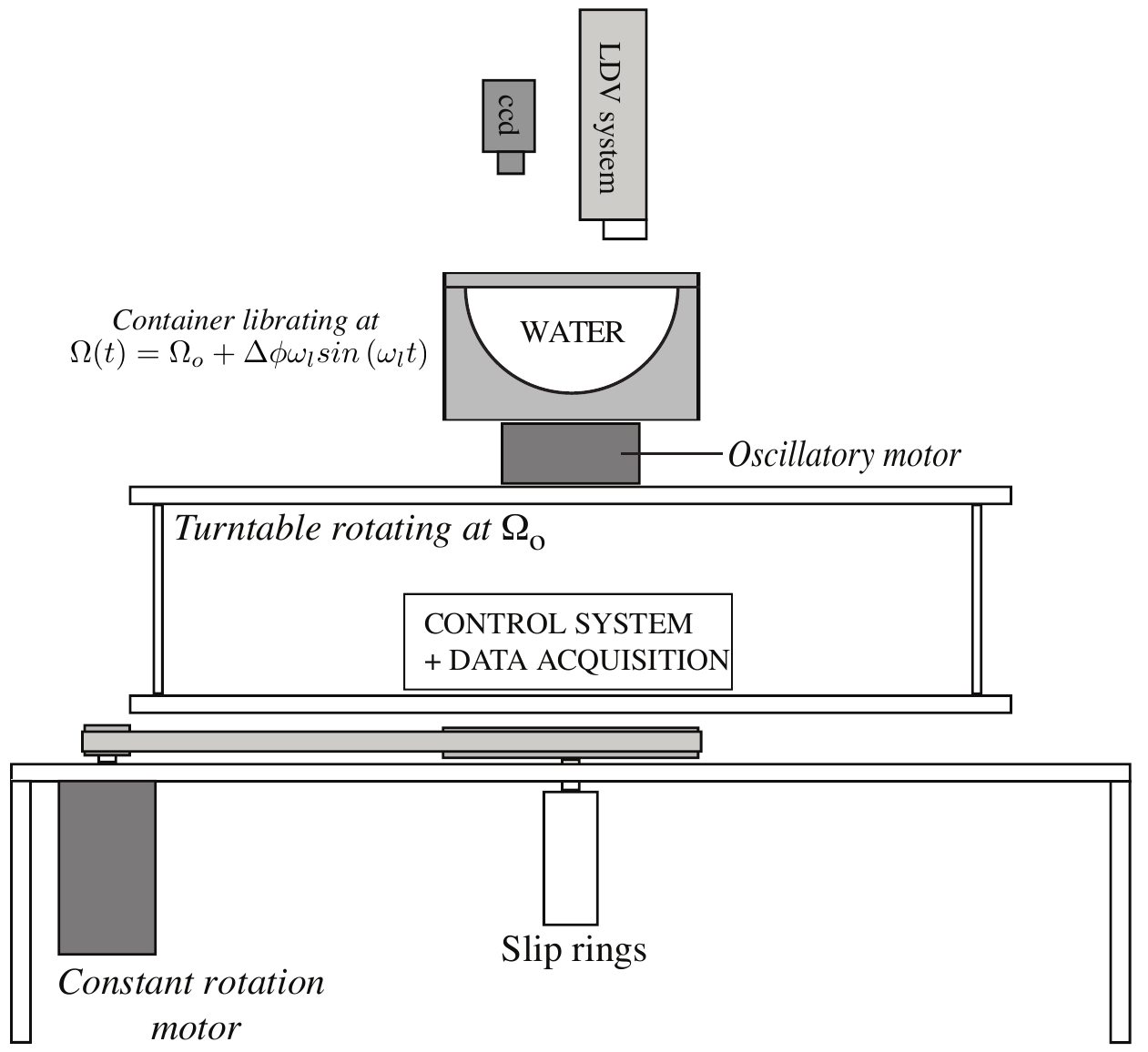}
    \caption{\it{Schematic view of the laboratory experiment and of the ultraLDV system
     \cite[][]{noir2010zf_a}, oriented to perform measurements of
the azimuthal velocity. No LDV measurements
of the azimuthal velocity can be readily performed in the case of an
ellipsoidal container with a non-axisymmetric equatorial
cross-section. Indeed, except if the LDV is moving with the tank, no
relative position of the LDV with the container can be found such
that the two laser beams remain coplanar over a libration cycle. A
co-librating configuration is not achievable in the present
experimental set-up due to the large centrifugal acceleration
involved. Therefore, LDV measurements are performed with a
"half-ellipsoidal" container, the northern half of the ellipsoid
being replaced with a flat plate of acrylic.}}
    \label{oblong_spheroid2}
  \end{center}
\end{figure}

To further validate the existence of LDEI, we have also performed
selected laboratory experiments (fig. \ref{oblong_spheroid2}). 
These allow us to reach
smaller values of the Ekman number, and hence to access more
chaotic flows, with the inconvenience being the difficulty in acquiring
precise quantitative measurements and in changing the shape of the
container. Except for the container, the laboratory
apparatus is the same as in ref. \cite{noir09} and \cite{noir2010zf_a}.
It consists in an oscillating tank filled with water and centered on
a turntable rotating at a constant angular velocity
$\Omega_d=0.5\, \textrm{Hz}$. In order to perform quantitative measurements
using Laser Doppler Velocimetry (LDV), the container consists in one
hemispheroid CNC (Computer Numerical Control) machined from cast acrylic cylindrical blocks that
is polished optically clear, with a top flat lid that avoids optical
distortions. This geometry, corresponding to a ``half-ellipsoidal'' 
container, does not allow modes of the elliptical instability that are 
antisymmetric around the equator. Note also that because of manufacturing constraints,
the small axis and the rotation axis of the container are equal,
with $a=127\ \textrm{mm}$, $b=c=89\ \textrm{mm}$. The experimental
parameters are then $\beta=0.34$, $c^*=0.812$ and $E=2.7 \times
10^{-5}$; $f$ and $\varepsilon$ have been changed systematically to
explore the ranges $f \in [0.5-2]$ with $\varepsilon=1$, and
$\varepsilon \in [0-1.6]$ with $f=1$, respectively. 
Similar to the numerical experiments, 
we observe a resonance band, here $f \in [1.43-1.66]$, that is 
characterized by strong, intermittent, space-filling turbulence. These cases are
marked by periods dominated by strong, small-scale, shear structures, followed by relaminarization period. These flows are not due to shear instabilities since the system does not exhibit any turbulence at higher frequency, for which the Rossby number is larger. Since no direct resonance can occur, if a shear instability were developing for a critical Rossby number it should remain unstable at larger forcing, regardless of the frequency. 

\begin{figure}
  \begin{center}
    \leavevmode
    \includegraphics[scale=0.3]{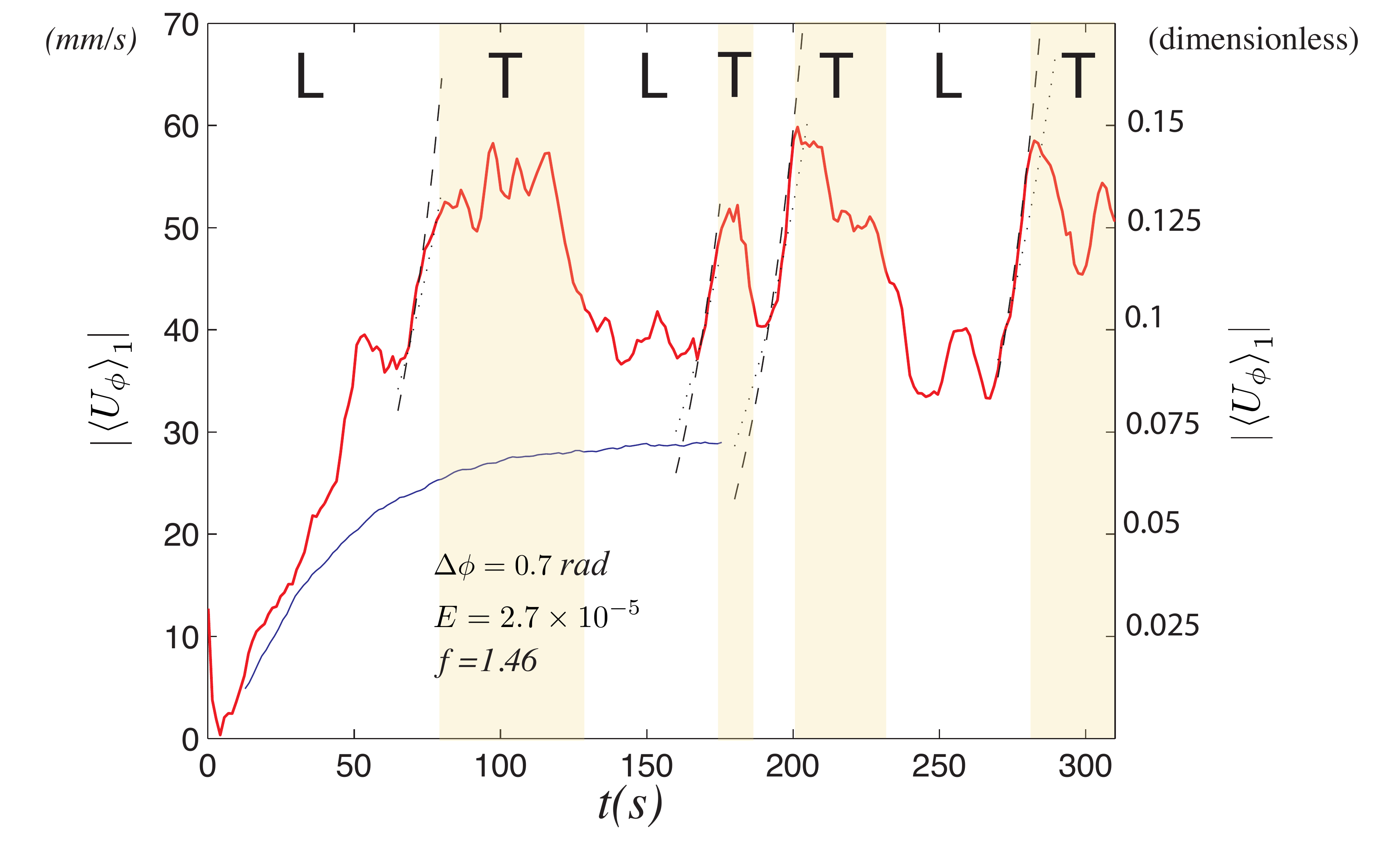}
    \caption{\it{(color online) Time evolution of
the norm of the azimuthal velocity averaged over 10 oscillations for $\Delta\phi=0.7\ \textrm{rad}$ ($\varepsilon
\sim 1$) with $\beta=0.34$ and $f=1.46$ (continuous upper curve, red online); $\beta=0.06$ and $f=1.40$  (continuous upper curve, blue online). The measurements are performed at a cylindrical radius $S_i=48\ \textrm{mm}$ along the short axis of the mean equatorial ellipse, 1 cm below the top flat surface. We perform a sliding window averaging over $10$ oscillations with an overlap of 90$\%$. In addition we represent the WKB exponential growth for to two extreme values of the damping factor, $K=1$ (dotted black) and $K=10$ (dashed black). The letters L and T stand for Laminar and Turbulent. The periods of turbulence, as observed in direct visualizations, are qualitatively represented by the bands.}}
    \label{ZF_30RPM_073HZ_80DEG_LDA_VIDEO}
  \end{center}
\end{figure}

An example of the
measured azimuthal velocity is shown in figure \ref{ZF_30RPM_073HZ_80DEG_LDA_VIDEO}
for $f=1.46$. As mentioned earlier, the observed chaotic
behavior of the flow as well as the fact that this behavior only
appears for specific frequencies are characteristic of the
elliptical instability. Assuming the existence of a resonant peak of
LDEI at $f=1.46$, the WKB approach (\ref{sigmath}) yields a typical
growth time $1/(\sigma\Omega_d)$ of the instability ranging between
$21s$ for a lower bound of the damping factor $K=1$ and $37s$
for an upper bound of the damping factor $K=10$. As shown in
figure \ref{ZF_30RPM_073HZ_80DEG_LDA_VIDEO}, these values are in good agreement with measurements during each growing phase of
the azimuthal flow, further validating the interpretation of this
chaotic behavior in terms of LDEI. Finally, note that even if no resonant cases have been obtained in this work with $f \leq 1$ \cite[the relevant planetary range for $f$ according to ref.][]{noir09}, the WKB analysis predicts unambiguously that the LDEI can be excited for an arbitrary $|f|<4$ provided that E is sufficiently small. Experiments at lower E should confirm it in the future. But the first
experimental results presented here, in addition to the numerical simulations
and in agreement with the theoretical analysis, already illustrate the
generic feature of the libration driven elliptical instability, which appears
for different geometries and various ranges of parameters.

In conclusion, studies of flows driven by longitudinal libration made using a spherical
CMB approximation suggest that purely viscous coupling does not lead to
significant planetary energy dissipation, angular momentum transfer nor
magnetic field induction \cite[][]{calkins2010lib}. These conclusions should be re-addressed in
accounting for the specific triaxial shape of the considered
planets, since space-filling
turbulence is observed in the present numerical and laboratory experiments in which
LDEI is excited. A complete understanding of the elliptical
instability excited by libration in celestial bodies requires the
characterization of the inertial modes, their frequency and their
viscous decay factor in the appropriate geometry and for the low values of
the Ekman number relevant to planetary applications. Nevertheless, the relevance of a LDEI mechanism
at planetary settings can be ascertained using the theoretical WKB
approach presented here, by estimating a lower bound of the
equatorial ellipticity leading to a positive growth rate $\sigma$.
Using typical values for forced longitudinal libration \cite[][]{noir09} of $f=1$, $E \sim
10^{-14}$, $\varepsilon \sim 10^{-4}$, a decay factor $K=1$ and
assuming a perfect resonance, equation (\ref{sigmath}) yields to a minimum equatorial ellipticity $\beta \sim 10^{-3}$ for excitation of LDEI. This first order
estimate is qualitatively comparable with the values expected for
Mercury, Europa or Io. Thus, the space-filling turbulence resulting from LDEI should exist within the fluid interiors of librating terrestrial bodies.\\
\newpage

For this work, D. C\'ebron was partially supported by the ETH Z\"urich Postdoctoral fellowship Progam as well as by the Marie Curie Actions for People COFUND Program.

\end{document}